\baselineskip=18pt  
\magnification=\magstep1  
\def\gtorder{\mathrel{\raise.3ex\hbox{$>$}\mkern-14mu  
             \lower0.6ex\hbox{$\sim$}}}  
\def\ltorder{\mathrel{\raise.3ex\hbox{$<$}\mkern-14mu  
             \lower0.6ex\hbox{$\sim$}}}  
\def\proptwid{\mathrel{\raise.3ex\hbox{$\propto$}\mkern-14mu  
             \lower0.6ex\hbox{$\sim$}}}  
  
\font\maintitle= cmbx10 at 15.00truept  
\font\sechead= cmbx10 at 12.00truept   
  
\def\V#1{{\bf #1}}  
\def\pa{\partial}  
\def\na{\nabla}  
\def\ti{\times}  
\def\bphi{B_\phi}

\centerline{\maintitle INSTABILITY OF TOROIDAL MAGNETIC FIELD}  
  
\bigskip  
\centerline{\maintitle IN JETS AND PLERIONS}  
  
\bigskip  
\medskip  
\centerline{Mitchell C. Begelman\footnote{$^1$}{Department of Astrophysical and  
Planetary Sciences, University of Colorado, Boulder, CO  80309}$^,$\footnote{$^2$}{I: mitch@jila.colorado.edu}}  
 
\medskip  
\centerline{JILA, University of Colorado and National Institute of Standards and Technology,}  
\centerline{Boulder, CO 80309--0440} 
 
\bigskip  
\centerline{\it Submitted to Ap.~J., April 28, 1997; in revised form, July
16, 1997; accepted August 2 1997.}  
\bigskip  
  
\centerline{\sechead ABSTRACT}  
  
\medskip  
Astrophysical jets and pulsar-fed supernova remnants (plerions) are expected to develop highly  
organized magnetic structures dominated by concentric loops of toroidal field, $\bphi$.  It has  
been argued that such structures could explain the polarization properties of some jets, and  
contribute to their lateral confinement through magnetic tension forces. A concentric toroidal  
field geometry is also central to the Rees--Gunn model for the Crab Nebula --- the archetypal  
plerion --- and leads to the deduction that the Crab pulsar's wind must have a weak magnetic  
field.  Yet this kind of equilibrium between magnetic and gas pressure forces, the ``equilibrium  
Z--pinch" of the controlled fusion literature, is well known to be susceptible to disruptive  
localized instabilities, even when the magnetic field is weak and/or boundary conditions (e.g., a  
dense external medium) slow or suppress global modes.  Thus, the magnetic field structures  
imputed to the interiors of jets and plerions are unlikely to persist for very long.  
  
To determine the growth rates of Z--pinch instabilities under astrophysical conditions, I derive  
a dispersion relation that is valid for the relativistic fluids of which jets and plerions may be  
composed, in the ideal magnetohydrodynamics (MHD) limit.  The dominant instabilities are  
kink ($m=1$) and pinch ($m=0$) modes. The former generally dominate, destroying the 
concentric field structure and probably driving the system toward a more chaotic state in which 
the mean field strength is independent of radius (and  in which resistive dissipation of the field 
may be enhanced). I estimate the timescales over which the field structure is likely to be 
rearranged and relate these to distances along relativistic jets and radii from the central pulsar in 
a plerion.  
  
I conclude that the central tenet of the Rees--Gunn model for the Crab Nebula --- the existence  
of a concentric toroidal field well outside the pulsar wind's termination shock --- is physically  
unrealistic.  With this assumption gone, there is no dynamical reason to conclude that the 
magnetic energy flux carried by the pulsar wind is much weaker than the kinetic energy flux.  
Abandoning the principal conclusion of Rees \& Gunn would resolve a long-standing puzzle in 
pulsar wind theory. 
  
\bigskip  
  
\centerline{\sechead 1.  INTRODUCTION}  
  
\medskip  
  
The poloidal (i.e., radial: $B_z$) and toroidal ($\bphi$) components of magnetic field frozen  
into a perfectly conducting outflow behave differently as the flow expands. If the fluid in a jet  
with cylindrical radius $r(z) \ll z $ and speed $v(z)$ expands laterally as it moves outward  
along  the $z$--axis, the poloidal field varies as $r^{-2}$ while the toroidal field varies as $r^{- 
1} v^{-1}$ (Begelman, Blandford, \& Rees 1984). Thus the toroidal field should eventually  
become dominant. The toroidal component of the field is expected to be even more dominant  
in the wind from a pulsar, since it is amplified initially by the pulsar's spin.  In the wind from the  
Crab pulsar, for example, the toroidal and poloidal field components should be comparable  
near the light cylinder radius, $2 \times 10^8$ cm.  By the time the wind reaches the reverse  
shock (and inner boundary of the visible nebula), at about $10^{17}$ cm, $\bphi$ is expected  
to be nearly 9 orders of magnitude stronger than $B_z$.   
  
Theorists have made much of the possible astrophysical importance of organized field  
structures dominated by concentric loops of toroidal magnetic field.  First, such field structures  
can have important dynamical consequences, for example, providing substantial pressure  
confinement in a jet (Benford 1978; Chan \& Henriksen 1980; Bicknell \& Henriksen 1980; Eichler 1993), via magnetic tension.  Second, the presumed existence of  
a  highly organized field has been used to place tight constraints on otherwise unobservable  
properties of the Crab pulsar's wind.  By assuming a concentric toroidal field in the Crab  
Nebula, Rees \& Gunn (1974) were able to deduce that the kinetic energy flux in the wind must  
be 2--3 orders of magnitude larger than the Poynting flux, at the radius at which the wind  
encounters the reverse shock.  This empirical conclusion has been confirmed and refined by  
later analyses (Kennel \& Coroniti 1984a; Emmering \& Chevalier 1987; Begelman \& Li 
1992) but has proven hard to reconcile with theoretical models of pulsar winds, which predict 
comparable fluxes of kinetic and electromagnetic energy (Michel 1969; Kennel, Fujimura, \& 
Okamoto 1983; Coroniti 1990; Arons 1992; Begelman \& Li 1994; Tomimatsu 1994; Michel 
1994; Melatos \& Melrose 1996). 
 
The applications cited above depend critically on the assumption that the field lines remain 
concentric within the Crab Nebula, or within an expanding  jet. But there are good reasons to 
question the plausibility of this assumption.  Plasma physicists have known for more than 40 
years that simple field configurations consisting of concentric circular loops of  
field are generically unstable, even when the magnetic pressure is weak compared to the gas  
pressure (Kruskal \& Schwarzschild 1954; Kadomtsev 1966). Such systems, called  
``Z--pinches" in the plasma physics literature (because the current flows in the $z-$direction), 
are subject to short-wavelength modes that cannot be suppressed by a judicious choice of 
boundary conditions, such as a very dense external medium or a highly supersonic flow velocity 
(e.g., K\"onigl \& Choudhuri 1985). Even  if the poloidal field were amplified by shear, as 
seems to occur in some jets, the poloidal field structure would lack the large-scale coherence  
--- in contrast to $\bphi$ --- necessary to suppress kink ($m=1$) and pinch ($m=0$) 
instabilities through tension forces. In particular, modes with an $m=1$ angular dependence 
(i.e., $\propto e^{i\phi}$) will destroy the concentric symmetry, leading to a much less regular 
field structure. 
 
In this paper, I discuss the effects of short-wavelength Z--pinch instabilities as they apply to  
astrophysical jets and pulsar-fed nebulae (plerions) such as the Crab Nebula.  By ``short"  
wavelength, I mean wavelengths that are much shorter than any of the length scales associated  
with the unperturbed medium, such as the width of the jet or radius of the plerion.  These modes are driven by forces due to the curvature of field lines and magnetic and gas pressure gradients in the background equilibrium.  They are distinct from the Kelvin-Helmholtz modes, driven by shear in the velocity field, which have been studied extensively in connection with jets.  These local modes are also distinct from the global pinch and kink modes that can affect magnetized jets under certain conditions (Chan \& Henriksen 1980; Eichler 1993).  I estimate the growth rates of the most unstable Z--pinch modes, and assess how these modes might reorganize the structure of the magnetic field, given the available evolutionary timescales. 
 
I conclude that, while the magnetic field in the Crab Nebula and many jets may remain aligned  
with the toroidal plane, the large-scale concentric structure of the field is probably destroyed by  
instabilities. This conclusion, if true, would obviate Rees \& Gunn's dynamical argument that 
the kinetic energy flux must exceed the Poynting flux in the Crab pulsar wind. 
  
The plan of this paper is as follows. Since the properties of Z--pinch instabilities seem not to be  
very familiar to the astrophysical community, I present heuristic derivations of the $m=0$ and 
$m=1$ stability criteria in \S~2.  Readers may also wish to consult standard textbooks 
(especially Bateman 1978; also Krall \& Trivelpiece 1973; Chen 1974; Jackson 1975) and 
monographs (e.g., Freidberg 1982) for further background on these well-studied instabilities, 
particularly in the context of controlled fusion devices.  The full local (i.e., short-wavelength) 
dispersion relation is derived in \S~3, taking special relativistic effects into account. I am not 
aware of other fully relativistic calculations of the Z--pinch dispersion relation in the literature.  
(In deriving the dispersion relation, I include perturbations of a small, uniform $B_z$. This can 
either enhance or detract from stability, depending on the helicity of the mode.)  Important 
consequences of the instability, notably growth rates and nonlinear development, are discussed 
in \S~4.  Section 5 deals with applications to jets and the Crab Nebula.  I summarize the main 
results in \S~6.    
    
\bigskip  
  
\centerline{\sechead 2. HEURISTIC DERIVATION OF STABILITY THRESHOLD}  
  
\medskip  
  
Consider a cylindrically symmetric MHD equilibrium with a magnetic field of the form  
$$ {\bf B} = B(r)\hat\phi +  b \hat z , \eqno(2.1) $$  
where $b$ is a constant $\ll B(r) $.  We assume that $b$ is independent of $r$ so that it does  
not figure in the equilibrium condition,  
$$  {\pa p\over \pa r} + {1\over 8 \pi r^2} {\pa (r^2 B^2) \over \pa r} = 0, \eqno(2.2) $$  
where $p$ is the gas pressure.  Despite its irrelevance to the equilibrium state, the inclusion of a  
small $B_z$ component has an important effect on stability.  It can either suppress or enhance  
instability at sufficiently short wavelengths, and dramatically affects the formal stability  
threshold for $m=0$ (axisymmetric) modes, as we will show below. Strictly speaking, a field  
configuration with the form (2.1) is a special case of  a ``screw pinch" (Freidberg 1982),  
because individual field lines are cylindrical helices; however, we will continue to use the term  
Z--pinch to express its principal characteristic.  
  
To provide some physical insight into a Z--pinch, we follow what happens when a narrow ring  
of magnetic flux is displaced slightly against an unperturbed background. For ease of  
visualization, we restrict our analysis to $m=1$ (kink) and $m=0$ (pinch) modes.  
  
Consider first an $m=1$ displacement, which corresponds to moving a flux loop sideways in  
the $r-\phi$ plane (neglecting the helical rise) without changing its radius or circular shape.  If  
we make this change slowly enough (simulating conditions near marginal stability), then the gas  
pressure will remain uniform around the loop; since the volume of the loop is unchanged to  
first order, the gas pressure is also unchanged,   
$$p_1' = p_0 . \eqno(2.3)$$  
We denote conditions at the unperturbed location by subscript 0, conditions at a perturbed  
location by subscript 1, and distinguish conditions inside the perturbed ring from those on the  
outside by a prime.  Now concentrate on a portion of the loop that moves from $r_0$ to $r_1  
= r_0 + \Delta r $.  Since the change is slow, hydrostatic equilibrium in the $z-$direction is  
maintained locally as a function of  $r$, implying  
$$ {\pa \over \pa r} \left[ { (B_1')^2\over 8\pi } + p_1'  \right] = {\pa \over \pa r}   
\left[ { (B_1)^2\over 8\pi } + p_1 \right] = - { B_1^2 \over 4 \pi r_1} ,  \eqno(2.4) $$  
where the final equality comes from eq.~(2.2).  Moreover, if we integrate eq.~(2.4) over radius  
the constant of integration must vanish so that   
$$   {(B_1')^2\over 8\pi } + p_1'  =  { (B_1)^2\over 8\pi } + p_1 . \eqno(2.5) $$  

In general, hydrostatic equilibrium cannot be satisfied simultaneously in both the $r-$ and  
$z-$ directions.  There will either be a restoring force in the $r-$direction, which moves the 
loop back toward its original location, or a destabilizing force which augments the 
perturbation. When $b \neq 0$, one must include the magnetic tension force due to the 
perturbed $B_z$.  For $m=1$ perturbations that are periodic in the $z-$direction with 
wavenumber $k$, the force due to the perturbed $B_z$ can be written as the sum of two 
terms, $-(2 kbB/r + k^2 b^2)  \Delta r /  
4\pi$.  The quadratic term ($\propto (kb)^2$) results from the distortion of $B_z$ into $B_r$,  
giving rise to a tension force in the $r-z$ plane which always acts as a restoring force. The term  
linear in $kb$ results from the distortion of $B_z$ into a combination of $B_r$ and $B_\phi$.  
The resulting tension force (in the $r-\phi$ plane) can either stabilize or destabilize the  
perturbation, depending on whether $kb$ is positive or negative.  If we visualize the $m=1$  
mode as a helix wound about the $z-$axis, then $kb > 0 $ corresponds to a helix wound in the  
same sense as the magnetic field.  The increased twist in $B_z$ then increases the tension in the  
$r-\phi$ plane, yielding a stabilizing effect. Conversely, a mode with $kb < 0$ is wound counter  
to the magnetic field.  The twist in $B_z$ then decreases the net tension, helping to destabilize  
the mode.  
  
The condition for stability is  
$$ \left\{ {\pa \over \pa r} \left[ { (B_1')^2\over 8\pi } + p_1'  \right] +  
{ (B_1')^2 \over 4 \pi r_0} + (2 kbB +  k^2 r b^2) {\Delta r \over  4\pi r_0 } \right\} \Delta r >   
0 . \eqno(2.6) $$  
We use $r_0$ in the term for the perturbed magnetic hoop stress because the radius of the loop 
is unchanged.  Eliminating the partial derivative through the use of eq.~(2.4) and noting that   
$$ { (B_1')^2 \over 4 \pi r_0} - { B_1^2 \over 4 \pi r_1} = \left({ B^2 \over 4 \pi r^2}  +   
  {2\over r}  {\pa  p\over \pa r} \right) \Delta r  \eqno(2.7) $$  
where we have used eqs.~(2.5) and (2.3) (and have dropped subscripts as permitted, to first  
order), we obtain    
$$ \left[ { B^2 \over r}  + 8 \pi   {\pa  p\over \pa r}   +  ( 2 k b B + k^2 r  b^2 ) \right]  (\Delta  
r)^2 >  0   \eqno(2.8)  $$  
as the condition for stability .  Substituting for the pressure gradient from eq.~(2.2) and noting  
that $(\Delta r)^2$ is positive-definite, we finally obtain the necessary condition for stability of  
$m=1$ modes:  
$$ { d \ln B \over d \ln r } < - {1 \over 2} + kr {b\over B} +   
{1 \over 2} \left( kr {b\over B}\right)^2  . \eqno(2.9) $$  
Thus, any region of a Z--pinch in which $B$ increases with radius or decreases more slowly  
than $r^{-1/2}$ should be subject to $m=1$ instability for $1 \ll kr \ll B/b $. We verify this  
result using the full dispersion relation in \S~3.  
  
Next consider $m=0$ modes, which correspond to changing the radius of a flux loop (from  
$r_0$ to $r_1 = r_0 + \Delta r$)  while keeping its center fixed.  Pressure balance in the $z- 
$direction again requires that eqs.~(2.4) and (2.5) be satisfied.   The stability condition is then    
$$\left\{ {\pa \over \pa r} \left[ { (B_1')^2\over 8\pi } + p_1'  \right] +  
{ (B_1')^2 \over 4 \pi r_1} + {k^2 r b^2 \over 4\pi r} {\Delta r \over r}  \right\} \Delta r  > 0 .  
\eqno(2.10) $$  
Note that we now use $r_1$ in the hoop stress term (instead of $r_0$) because the radius of  
the loop has changed, and that only the quadratic term in the $B_z$ tension force appears for  
$m=0$ perturbations.  Using eqs.~(2.2) and (2.4) and defining $\Delta B \equiv B_1' - B_0 $,  
we can manipulate eq.~(2.10) into the form  
$$  { d \ln B \over d\ln r } <    {\Delta B / B \over \Delta r / r } + {1 \over 2} \left( kr {b\over  
B}\right)^2  . \eqno(2.11) $$  

To evaluate $(\Delta B / B) / (\Delta r / r)$, we define $\Delta p \equiv p_1' - p_0$ and use  
eqs.~ (2.2) and (2.5) to show that   
$$ 4\pi \Delta p = -  B^2 \left( {\Delta B \over B} + {\Delta r \over r} \right) . \eqno(2.12) $$   
We must then consider the cases of finite and vanishing $b$ separately.  If $b = 0$ identically,  
then flux loops are truly closed.  The gas pressure responds to changes in the volume of the  
loop, $V$, according to an adiabatic law,   
$$ {\Delta p \over p} = - \gamma {\Delta V \over V} = - \gamma \left(    
{\Delta A \over A} + {\Delta r \over r} \right), \eqno(2.13)  $$  
where $A$ is the cross-sectional area along the loop.  From magnetic flux conservation, we  
also have  
$$ {\Delta B \over B} = - {\Delta A \over A} . \eqno(2.14) $$  
Combining eqs.~(2.12), (2.13), and (2.14) with eq.~(2.11), and adopting the usual plasma  
`beta' parameter $\beta \equiv  8\pi p / B^2$, we obtain the stability criterion  
$$  { d \ln B \over d\ln r } <    { \gamma\beta  - 2 \over \gamma\beta + 2} . \eqno(2.15) $$  

Surprisingly, the stability threshold with $b \neq 0$ is quite different from eq.~(2.15), even if  
$b$ is vanishingly small.  In this case, the flux loops are connected to form a cylindrical helix.   
Near marginal stability, fluctuations are extremely slow so there is time for the pressure to  
equilibrate along the helix.  Therefore $\Delta p = 0$, eq.~(2.12) gives $(\Delta B / B) / (\Delta  
r / r) = -1$, and the stability criterion follows immediately from eq.~(2.11).  In \S~3 we verify  
both cases of the $m=0$ stability threshold, and assess their physical implications in later  
sections.  
  
\bigskip  
  
\centerline{\sechead 3. DERIVATION OF DISPERSION RELATION}  
  
\medskip  
\centerline{3.1. {\it Linearized Equations} }  
  
\medskip  
  
The relativistic momentum equation for a magnetofluid can be written in the form  
$$ \Gamma^2 \left(\rho + {p\over c^2} \right) \left( {\pa \V v \over \pa t} + \V v \cdot \na \V  
v \right) + \na p + {\V v \over c^2} {\pa p \over \pa t} - \rho_e \V E - {\V j \ti \V B \over  
c} = 0  \eqno(3.1) $$  
(e.g., Weinberg 1972), where $\Gamma^2 \equiv (1 - v^2 /c^2)^{-1} $, $p$ and $\rho$ are the  
proper pressure and total energy density, respectively, $\rho_e$ is the charge density, and the  
electromagnetic fields and current density are represented by their usual symbols.  In this paper  
we are concerned with perturbations about time-independent, static equilibria, and therefore we  
treat the velocity $\V v$ as a first-order quantity and $\Gamma-1$ as second-order.  By the  
usual arguments of MHD, the electric field $\V E$ is first-order, and since $\rho_e = \na\cdot  
\V E / 4\pi $ the electrostatic force term is second-order.  Distinguishing zeroth from   
first-order quantities by the corresponding subscripts, we obtain the following first-order  
momentum equation:  
$$ \left(\rho_0 + {p_0\over c^2} \right) {\pa \V v \over \pa t} = - \na p_1 + {\V j_1 \ti \V  
B_0 \over c} + {\V j_0 \ti \V B_1 \over c},  \eqno(3.2) $$  
which must be solved subject to the perturbed Maxwell's equations:  
$$ \eqalignno{  
\na\ti \V B_1 &= {1\over c} {\pa \V E_1\over \pa t} + {4\pi \V j_1 \over c} &(3.3) \cr  
\na\ti \V E_1 &= - {1\over c} {\pa \V B_1\over \pa t} &(3.4) \cr  
\na \cdot \V B_1 &= 0; &(3.5) \cr  }$$   
Ohm's law:  
$$ \V E_1 + { \V v \ti \V B_0 \over c} = 0; \eqno(3.6) $$  
and an equation of state, which we take to be  
$$ {\pa p_1 \over \pa t} = - (\V v \cdot \na) p_0 - \gamma p_0 (\na \cdot \V v), \eqno(3.7)$$  
\noindent where $\gamma$ is the adiabatic index.  Note in equation (3.3)  
that we retain the displacement current $\pa \V E_1 / \pa t$, which is usually dropped in  
nonrelativistic MHD analyses.  
  
Using eq.~(3.6) to eliminate $\V E_1$ from equations (3.3) and (3.4), we obtain  
$$ \V j_1 = {c\over 4\pi} (\na\ti \V B_1 ) + {1 \over 4\pi c} \left[ \left( {\pa \V v \over \pa  
t } \right) \ti \V B_0 \right],  \eqno(3.8) $$  
which includes the displacement current, and   
$${\pa \V B_1 \over \pa t} = \na \ti (\V v \ti \V B_0), \eqno(3.9) $$  
the usual flux-freezing equation.  We now substitute for the current densities in  
eq.~(3.2) using eq.~(3.8) for $\V j_1$ and $4\pi\V j_0 / c  = \na \ti \V B_0$, obtaining (after  
some manipulation)  
$$\eqalignno{   
{\V j_1 \ti \V B_0 \over c} + {\V j_0 \ti \V B_1 \over c} &= {1\over 4\pi} \left[ (\V B_1 \cdot  
\na) \V B_0 + (\V B_0 \cdot \na) \V B_1 - \na (\V B_0 \cdot \V B_1) \right] \cr  
  &-  {1\over 4\pi c^2} \left[ B_0^2 {\pa\V v \over \pa t} - \V B_0 \left( \V B_0 \cdot {\pa\V  
v \over \pa t} \right)  \right] . &(3.10) \cr   }$$  
Next we substitute eq.~(3.10) into eq.~(3.2), differentiate the entire equation with respect to  
time, and substitute eq.~(3.7) to eliminate $p_1$.  After some rearrangement of terms, the  
resulting equation can be written  
$$ \eqalignno{  
\biggl( \rho_0 + {p_0\over c^2} &+ {B_0^2 \over 4\pi c^2}  \biggr) {\pa^2 \V v \over \pa t^2}  
- { \V B_0 \over 4\pi c^2} \left( \V B_0 \cdot {\pa^2 \V v \over \pa t^2 } \right)   
= \na [ (\V v \cdot \na ) p_0 +  \gamma p_0 (\na \cdot \V v) ] \cr  
&+ {1\over 4\pi} \left[ \left( {\pa \V B_1 \over \pa t} \cdot \na\right) \V B_0   
+ (\V B_0 \cdot \na) {\pa \V B_1 \over \pa t} - \na \left(\V B_0 \cdot {\pa \V B_1 \over \pa  
t} \right) \right] . &(3.11)   \cr   }$$  
Note that the displacement current gives rise to an effective inertia associated with the  
magnetic energy density, for velocity perturbations which are transverse to the unperturbed  
magnetic field.  One can derive a homogeneous linear equation in $\V v$ by substituting for  
$\pa \V B_1/ \pa t$ from eq.~(3.9), but before taking this step we will first adopt some  
simplifying assumptions about the nature of the underlying equilibrium and of the perturbations  
of interest.     
  
\medskip  
  
\centerline{3.2. {\it Quasi-Local Perturbations About Toroidal--Field Equilibria}}  
  
\medskip  
  
As in \S~2, we specialize to perturbations about cylindrically symmetric equilibria (in  
coordinates $\{r, z, \phi\}$) satisfying   
$$ p_0 = p_0 (r), \qquad \V B = B_0 (r) \hat \phi  + b \hat z , \eqno(3.12) $$   
where $b $ is a constant. Such equilibria must satisfy eq.~(2.2).  We further specialize to   
short-wavelength (quasi-local) modes for which all perturbed quantities can be approximated  
to have the form   
$$ e^{i(m\phi + kz  + lr - \omega t)} .  \eqno(3.13) $$   
The main requirement for such modes (assuming $d\ln B / d\ln r \sim O(1)$) is that $k^2r^2,  
l^2r^2 \gg 1$.  Since the background equilibrium depends on radius, true normal modes of the  
system have the more general form $ f(r) e^{i(m\phi + kz - \omega t)}$. To check the   
self-consistency of our approximation scheme, we derived the exact global dispersion relation  
for the case $b = 0$, and showed explicitly that the exact equations reduce to our approximate  
equations under the assumptions adopted below.    
  
The quasi-local equations admit two classes of modes: ``high-frequency'' modes, the relativistic 
versions of magnetosonic and Alfv\'en waves, and ``low-frequency" modes, which include the 
instabilities we wish to study.  To select the latter, and to determine which terms to retain in the 
quasi-local approximation, we adopt to following orderings, the self-consistency of which we 
have verified a posteriori:  
$$  \eqalignno{    
k^2r^2 &\gg 1 + m^2, \qquad  O(k) \sim O(l),  &(3.14) \cr    
| \omega^2|  &\ll k^2 c^2,  \qquad kv_z + l v_r \sim O\left(  {v_r \over r } \right), &(3.15) \cr  
b &\ll B_0, \qquad bl, bk  \sim O\left(  {B_0 \over r } \right). &(3.16)  \cr   } $$  
To simplify notation, we will henceforth drop the subscript on unperturbed quantities. We then  
define the relativistic enthalpy,  
$$ \varepsilon \equiv \rho c^2 + p , \eqno(3.17) $$  
and the following dimensionless parameters:  
$$ \nu \equiv {\varepsilon \over \gamma p} {\omega^2 r^2 \over c^2}, \qquad \alpha \equiv { r  
\over B} {d B \over d r},   
\qquad \eta \equiv kr {b \over B}, \qquad \beta \equiv {8\pi p \over B^2} . \eqno(3.18) $$  
  
Using the approximations and parameters defined above, the components of $\pa \V B_1 / \pa  
t$ in eq.~(3.9) are given to lowest order by   
$$  \eqalignno{  
\left( { \pa \V B_1 \over \pa t }\right)_r &= - {k \over l} \left( { \pa \V B_1 \over \pa t  
}\right)_z  =  i{B\over r}(m + \eta) v_r,  &(3.19)\cr  
\left ( { \pa \V B_1 \over \pa t }\right)_\phi &=  {B\over r} \left[ (1 - \alpha ) v_r  -   
r (\nabla\cdot \V  v ) +  i( m + \eta)  v_\phi \right] ,  &(3.20) \cr   }$$  
where we have used eq.~(3.15) to justify setting $v_z = -(l /k) v_r $ to lowest order in  
eq.~(3.19), and have substituted $k v_z + lv_r  = - i(\nabla\cdot \V v ) - m v_\phi/r  + iv_r / r$  
in eq.~(3.20).   
  
We now manipulate the momentum equation to derive the dispersion relation.  Writing  
eq.~(3.11) in the form  
$$ \V a = \nabla h  + \V F , \eqno(3.21) $$  
where $\V  a$ represents the acceleration terms and $h$ contains the gas pressure forces, we  
find the following expression for the components of the magnetic force, $\V F$, again keeping  
terms to lowest order:  
$$ \eqalignno{  
F_r &= {B\over 4 \pi r} \left\{  i\left[ m + \eta \left( 1 + {l^2\over k^2} \right) \right] \dot B_r  
- (\alpha + 1) \dot B_\phi -{\pa (r \dot B_\phi) \over \pa r}   \right\}  &(3.22) \cr  
F_z & =  - ikr{B\dot B_\phi \over 4 \pi r} &(3.23) \cr  
F_\phi &=  {B\over 4 \pi r} \left[  (\alpha + 1)\dot B_r  +  i\eta \dot B_\phi  \right]  , &(3.24)   
\cr }  $$  
where the dotted quantities represent the time derivatives (3.19) and (3.20).  
We will ultimately obtain the dispersion relation by evaluating the combination  
$$ \eqalignno{   
{\pa a_z\over \pa r} - {\pa a_r\over \pa z} &= ik \omega^2 \left( \rho + {p\over c^2} +   
{B^2 \over 4\pi c^2}  \right) \left( 1 + {l^2\over k^2} \right) v_r   \cr  
&={\pa F_z\over \pa r}- {\pa F_r \over \pa z} \cr  
&= ik {B\over 4 \pi r} \left\{   2 \dot B_\phi - i(m + \eta) \left( 1 + {l^2\over k^2} \right)  \dot  
B_r \right\}   &(3.25) \cr  }   $$  
where we have used the ordering adopted above to retain terms only to lowest nonvanishing  
order.  Note that the gas pressure terms cancel in this combination.   
Dividing through by $ikB^2/ (4\pi r^2)$, substituting for $\dot B_r$ from eq.~(3.19), and  
rearranging terms, we obtain  
$$ \left[  \left( {\gamma p \over \varepsilon} + {\gamma\beta  \over 2} \right)  \nu - ( m  + \eta  
)^2 \right]   
\left( 1 + {l^2\over k^2} \right) {v_r \over 2}  =  { r \dot B_\phi \over B }  \eqno(3.26) $$  
where $\nu$ is the dimensionless frequency defined in eq.~(3.18).  
  
We use the $z-$ and $\phi-$ components of eq.~(3.21) to eliminate $\dot B_\phi $ (or,  
implicitly, $\nabla \cdot \V v $ and $v_\phi$) in favor of $v_r$.  In the $z-$component, we  
may neglect the acceleration terms compared to the gas pressure and magnetic forces (just as,  
in our heuristic treatment [\S~2], we assumed magnetohydrostatic equilibrium in the vertical  
direction).  Using the equilibrium condition written in the form  
$$ {\pa  p \over \pa r} = - (\alpha + 1) {B^2 \over 4\pi r}  \eqno(3.27) $$  
to eliminate the pressure gradient, we find (after some algebra)  
$$ i(m+\eta) v_\phi = \left( 1 + {\gamma \beta \over 2} \right) r (\nabla\cdot \V v) - 2 v_r .  
\eqno(3.28) $$  
This allows the right-hand side of eq.~(3.26) to be simplified to  
$$ { r \dot B_\phi \over B }  =  {\gamma \beta \over 2} r (\nabla\cdot \V v) - (\alpha + 1)  v_r  
\eqno(3.29) $$  
Using eqs.~(3.27) and (3.28), the $\phi-$component of the momentum equation can be  
simplified to  
$$ - \nu  v_\phi   = i( m + \eta) r (\nabla\cdot \V v).  \eqno(3.30) $$  
Using eq.~(3.28) a second time to eliminate $v_\phi$, we obtain   
$$ \left[  (m +\eta)^2 - \nu \left( 1 + {\gamma \beta \over 2} \right)  \right] r (\nabla\cdot \V v)  
=  - 2  \nu  v_r . \eqno(3.31) $$  
Finally, eliminating $r (\nabla\cdot \V v)$ in eq.~(3.29), we are able to cancel factors of $v_r$  
in eq.~(3.26) to obtain the dispersion relation:  
$${1\over 2} \left[  \left( {\gamma p \over \varepsilon}  + {\gamma \beta \over 2}   \right)  \nu  
- ( m  + \eta )^2 \right] \left( 1 + {l^2\over k^2} \right) +  
{  (m+\eta)^2 - \nu \left( 1 - {\gamma \beta \over 2}\right)  
\over (m+\eta)^2 - \nu \left( 1 + {\gamma \beta \over 2}\right)} + \alpha = 0 .  \eqno(3.32) $$  
Equation (3.32) is the principal result of this section.  
  
Note that eq.~(3.32) can be written as a quadratic in $\nu$, $f(\nu) = A \nu^2 + B \nu + C =  
0$.  It is easy to show that $f(\nu)$ must change sign between its limit at $\nu \rightarrow  
\pm \infty$ and its value at $\nu = (m + \eta)^2 / (1 + \gamma\beta / 2) $. This implies that  
$B^2 - 4 AC \ge 0$, i.e., that $f(\nu)$ always has two real roots.  Since $\nu$ is effectively  
$\omega^2$, the modes are always either purely growing or purely damped as, of course, they  
must be for a nondissipative hydromagnetic system.     
  
\bigskip  
  
\goodbreak 
\centerline{\sechead 4. INTERPRETATION OF DISPERSION RELATION}  
  
\medskip  
\centerline{4.1. {\it Instability Threshold} }  
  
\medskip  
  
Unstable modes have imaginary $\omega$, implying $\nu < 0$.  We therefore seek values of  
$l^2 / k^2 > 0$, such that $f(\nu) = 0$ for $\nu < 0$.  Note that for $l^2 / k^2 \rightarrow  
\infty$, eq.~(3.32) implies that $\nu \rightarrow (m + \eta)^2 / ( \gamma p /  \varepsilon +  
\gamma\beta / 2 ) > 0$, so the modes are stable in this limit.  Unstable modes exist only for a  
finite range of   $l^2 / k^2$,   
$$  0 <  {l^2  \over  k^2}  <  2 {\alpha + 1 \over (m + \eta)^2 } - 1 , \eqno(4.1)$$  
where the latter portion of the inequality comes from setting $\nu = 0$ in  eq.~(3.32) with  
$(m+\eta)$ taken to be finite.  The necessary and sufficient condition for the existence of  
instability is then  
$$ \alpha  \equiv {d \ln B \over d\ln r} > {1\over 2} (m + \eta)^2 - 1, \eqno(4.2) $$  
which agrees exactly with the results of our heuristic analysis for both $m = 1$ (eq.~[2.9])  and  
$m = 0$ (eq.~[2.11] with $(\Delta B/B)/(\Delta r/r) = -1$). If $m + \eta = 0$ identically, the  
limit of  eq.~(3.32) as $\nu \rightarrow 0$ yields eq.~(2.15), also in agreement with our  
heuristic analysis for $m = b = 0$.    
  
\medskip  
\centerline{4.2. {\it Fastest Growing Modes} }  
  
\medskip  
  
The fastest growing modes are those with the most negative values of $\nu$.  Recalling that  
$\nu \equiv (\varepsilon / \gamma p) \omega^2 r^2 / c^2$, one sees from eq.~(3.32) that  
growth rates depend on the radial wavenumber, $l$, only through the squared ratio of radial to  
vertical wavenumber, $\delta \equiv l^2 / k^2$.  Growth rates also depend on $k$ through  
$\eta \equiv kr (b / B)$.    
  
As one might surmise from eq.~(4.1), the growth rates of unstable modes increase with  
decreasing $l$, a result which can be verified by differentiating eq.~(3.32) with respect to  
$\delta$, holding $k$ and $m$ fixed.  Defining $a\equiv (m + \eta)^2$,  $d \equiv \gamma\beta  
/ 2  $, and $e \equiv \gamma p / \varepsilon $, we find  
$$ \left[  {1 \over 2} (e + d) (1 + \delta) + { 2 a d\over [a - (1 + d)\nu ]^2 } \right]   
{\pa \nu \over \pa \delta } = {1 \over 2} [a - (e + d )\nu ]  . \eqno(4.3) $$  
The quantity in square brackets on the left-hand side is positive-definite for all $\nu$, while the  
right-hand side is positive-definite for $\nu < 0$.  Therefore, $\pa \nu / \pa \delta > 0$ for $\nu  
< 0$.  This means that the fastest growing modes occur for $\delta  \ll 1 $, i.e., for vertical  
wavelengths much shorter than radial wavelengths.  We will henceforth set $l^2 / k^2 \approx  
0$ in eq.~(3.32), which now takes the form  
$$ {1 \over 2} \left[ (e + d) \nu - a \right] + { 2d \nu \over a - (1 + d)\nu } + \alpha + 1 = 0  
. \eqno(4.4) $$  

Growth rates depend on $k$ and the mode number, $m$, through the combination $a = (m +  
\eta)^2$. Although $m$ must be an integer or zero, $a$ can take on any positive value, and  
unstable modes exist for $0 < a < 2 (\alpha + 1)$. Differentiating eq.~(4.4) with respect to $a$,  
we obtain  
$$ \left[{1 \over 2} (e + d) + { 2 a d\over [a - (1 + d)\nu ]^2 } \right]   
{\pa \nu \over \pa a} =   {1 \over 2} + { 2 d \nu \over [a -  (1 + d) \nu ]^2 } .  
\eqno(4.5) $$  
The quantity in square brackets on the left-hand side is positive-definite, but for $\nu < 0$ the  
right-hand side may be either positive or negative.  As $a$  approaches $2(\alpha + 1)$, $\nu$  
approaches zero and $\pa \nu / \pa a$ is positive.  This ensures that the growth rate increases  
with decreasing $a$ near the stability threshold.  As  $a$ decreases toward zero, two behaviors  
are possible:  1) $\nu$ is monotonic in $a$ with the largest growth rate at $a \rightarrow 0$, or  
2) $\nu$ exhibits a local minimum at finite $a$.  To see that there can be at most one local  
minimum and no local maxima, differentiate eq.~(4.5) with respect to $a$ and set $\pa \nu /   
\pa a = 0$, which shows that $\pa^2 \nu / \pa a^2 > 0$ for $\nu < 0$.   
  
To determine which of the two behaviors applies, evaluate eqs.~(4.4) and (4.5) at $a=0$ and  
eliminate $\nu$ between them.  If  $\pa \nu /  \pa a < 0 $ at $a=0$, then the maximum growth  
rate occurs at finite $a$.  This condition is equivalent to an upper limit on $\alpha$,  
$$ \alpha < {d - 1 \over d + 1} + { 2 d (e + d) \over (d + 1 )^2 }  . \eqno(4.6)   $$  
The inequality (4.6) is satisfied for a wide range of plausible equilibria, where: 1) $p$ decreases  
monotonically outward, and 2) the toroidal flux surfaces do not contain a neutral sheet (a  
surface where $B_\phi$ changes sign). In such equilibria, $\alpha > -1$ everywhere, and  
$\alpha \rightarrow -1$ for $d \rightarrow 0$ (low-$\beta$ limit).  If we approximate this class  
of equilibria by the analytic model 
$$ \alpha =  {\alpha_0 d - 1 \over d + 1} , \eqno(4.7)  $$  
then eq.~(4.6) is satisfied for $\alpha_0 < 3$.  Note that eq.~(4.7) with $\alpha_0 = 1$ is  
satisfied identically for purely toroidal MHD equilibria established by adiabatic flow with  
$v_\phi = 0$. These equilibria form a two-parameter family that has been studied by Begelman  
\& Li (1992) as a model for the evolution of the Crab Nebula, and may also be applicable to  
jets as they propagate and expand laterally.  Note also that models with $\alpha_0 < 1$ have  
divergent current density on the axis. 
  
To find the maximum growth rate for systems satisfying eq.~(4.6), we set the right-hand side  
of eq.~(4.5) equal to zero, and solve the resulting equation simultaneously with eq.~(4.4).  The  
minimum value of $\nu$,  
 $$ \nu_m = - {8d \over (1 - e)^2} \left\{ \left[  1 - { (\alpha + 1) (1 - e)  \over 4 d } \right]   
 - \left[  1 - { (\alpha + 1) (1 - e)  \over 2 d } \right]^{1/2}  \right\} ,  \eqno(4.8) $$ 
is attained at  
$$ a_m = \alpha + 1 + {1 \over 2} (e + 2d + 1) \nu_m  .  \eqno(4.9) $$ 
We chose the solution of the quadratic leading to eq.~(4.8) that gives positive $a_m$.   
 
\medskip  
  
\centerline{4.3. {\it Useful Approximations and Qualitative Summary} }  
  
\medskip  
 
The dispersion relation (4.4) is particularly simple to analyze in the limit of a weak magnetic  
field, $\beta \gg 1$ ($d \gg 1$).  To lowest order, the result is  
$$ \nu = d^{-1} \left[   1 + a -\alpha  - \sqrt{ (\alpha - 1)^2 + 4a } \right]  . \eqno(4.10) $$ 
Evaluating equations (4.8) and (4.9) in the weak-field limit, we find  
$$ \nu_m \approx - { (\alpha + 1) ^2 \over 4 d} \eqno(4.11) $$ 
with maximum instability occurring at  
$$  a_m \approx (\alpha + 1)  \left[  1 - {(\alpha + 1) \over 4 }  \right] . \eqno(4.12) $$ 
Equations (4.10)--(4.12) give remarkably accurate results over a much wider range of  
instability than the asymptotic regime where the approximations are rigorously justifiable, i.e.,  
where the field is weak.  Because of the structure of the full dispersion relation, eq.~(4.10)  
automatically gives the correct condition for $\nu$ to vanish, which occurs at rather small  
values of $\beta$ (e.g., at $\beta = 2/(3\gamma)$ for $a = \alpha_0 = 1$).  Thus, the  
approximation is ``pinned down" outside its nominal range of validity, a feature which could  
account for its usefulness. The accuracy of the approximation increases with increasing $a$ and  
decreasing $\alpha_0$. At $a = \alpha_0 = 1$, it is accurate to within about 15\% for all $d$  
(and gets better for higher $a$ and lower $\alpha_0$), but it is rather poor near the stability  
threshold, for $a \rightarrow 0$.   
 
Evaluation of the dispersion relation for different parameter combinations reveals other  
important trends.  $m =1$ modes (with $|\eta| \ll 1$)  are more unstable than $m=0$ modes in  
the weak-field limit, provided that $\alpha < 5/2$.  In regions where the field is relatively  
stronger ($\beta \ltorder 1$), $m= 0$ modes are relatively more important, and may begin to  
dominate for $\beta$ below some threshold value $\beta_{\rm th} > 2/(3\gamma)$.  For  
$\alpha_0$ close to 1, the  crossover in growth rates occurs close to the stability threshold for  
the $m=1$ mode.  As we have noted, in the special (but physically motivated) case $\alpha_0 =  
1$, there is no $m=0$  instability at all.  Note, however, that the $a=1$ modes are stable for  
$\alpha \le -1/2$, whereas instability of $a =0$ modes can persist until $\alpha \rightarrow -1$.   
 
Despite the relationship of $m=1$ and $m=0$  growth rates, there is no trend toward  
increasing instability with higher values of $a$ (or $m$).  Indeed, eq.~(4.12) implies that the  
fastest growing modes in the weak-field limit always occur for $a \le 1$. 
 
Finally, we note that the parameter $e$, which ranges between $(c_s / c)^2 \ll 1 $ in the  
nonrelativistic limit (where $c_s$ is the speed of adiabatic sound waves) and $1/3$ in the  
ultrarelativistic limit, has relatively little effect on the growth rate for plausible values of $d$,  
$a$, or $\alpha_0$. 
 
To summarize, the toroidal-field equilibria under consideration are most unstable to $m=1$ and  
$m=0$ modes. $m=1$ modes dominate in the weak-field limit when $\alpha_0 < 5/2$; when  
$\alpha_0$ is close to 1, they dominate throughout. (Of course, $m\ge 1$ modes can behave  
like lower-$m$ modes if $\eta$ is negative with a large absolute value, but we will not discuss  
this possibility here.)  For our purposes, eq.~(4.10) or (4.11) will suffice to give rough  
estimates of the growth rate for the most unstable modes.  Setting $\alpha = 1$ and restoring  
dimensions, we have 
$$ \omega^2 \sim - {c^2 \over r^2} {B^2 \over 4\pi \varepsilon}  . \eqno(4.13)  $$ 
The corresponding growth timescale is 
$$  t_g  = | \omega|^{-1} \sim  \left(  2 \beta \right)^{1/2}   {r\over c} \eqno(4.14) $$ 
in the ultrarelativistic limit ($\varepsilon = 4 p $), and 
$$  t_g  = \sim {r\over v_A} ,\eqno(4.15) $$ 
where $v_A$ is the Alfv\'en speed, in the nonrelativistic limit. 
 
\medskip  
  
\centerline{4.3. {\it Nonlinear Development} }  
  
\medskip  
  
The growth timescales derived above are sufficiently short, compared to the time available in  
systems like the Crab Nebula and jets, that the nonlinear development of the instabilities must  
be considered.  Given the problem's complexity, a qualitative analysis will have to  
suffice.  Even at this level, it is clear that the $m=0$ and $m=1$ modes will  produce  
dramatically different effects. 
 
Consider the $m=0$ modes first, supposing that these grow faster than the $m=1$ modes.   
Their initial effects will be to create gradients of $\bphi$ in the $z-$direction, without  
destroying the concentric field topology.  Equilibrium generally cannot be established under  
such conditions, suggesting that reorganization of the field's radial structure will continue well  
into the nonlinear regime.  What is the likely outcome of this reorganization?   
 
If the flow is adiabatic (even if the adiabats are different on different streamlines), then the  
system should evolve toward the family of equilibria derived by Begelman \& Li (1992).  These  
equilibria satisfy eq.~(4.7) with $\alpha_0 = 1$, {\it and are neutrally stable to $m=\eta = 0$  
modes at all radii}. We therefore conjecture that $m = 0$ modes, where dominant, will  
rearrange the field in such a way as to approach marginal stability.  Once this happens, we need  
only consider the $m = 1$ modes, which will be unstable for all $\beta > 2 / (3\gamma)$ 
($=1/2 \ (2/5)$ for an ultrarelativistic (nonrelativistic) equation  of state).  (If $B_z \neq 0$, 
then $m=0$ modes will still be slightly unstable, with growth rates $\sim O(\eta)$; however, 
the phase space of rapidly growing modes will be greatly reduced compared to the $m=1$ 
modes, and the latter should dominate.) 
 
Our heuristic derivation of $m=1$ instability in \S~2 seems also consistent with continued  
growth into the nonlinear regime.  Recall that modes with larger radial wavelengths (smaller  
radial wavenumber $l$) are more unstable than modes with shorter radial wavelengths, for 
fixed vertical wavenumber $k$.  Thus, relatively large sideways excursions of narrow field 
loops are possible before strong nonlinearity sets in.  The geometry of the unstable motion is 
reminiscent of a stack of coins coming apart.   
 
Once the displacement of a loop of field approaches and exceeds its initial radius, the loop no 
longer encloses the axis of symmetry and the dynamics of the instability becomes qualitatively 
different. Conditions would then seem to favor a significant reduction in magnetic field energy 
through contraction and reconnection of flux loops. Free of the inhibiting effects of a 
concentric field structure, closed flux loops would tend to contract, adiabatically converting 
magnetic field energy into particle energy.  Neighboring flux loops might also reconnect, 
turning magnetic into particle energy non-adiabatically.  There seems to be no reason to expect 
a concentric field structure to be restored, or its destruction inhibited, by nonlinear effects.  
Modeling these effects is beyond the scope of this investigation.  However, we make the 
following conjectures based on the nature of the unstable modes and their probable 
development: 
 
\smallskip 
 
\item{1)} Regions of concentric toroidal field with $\beta > 2/ (3\gamma) $ are subject to  
$m=1$ instability.  This will include the axis and regions nearby, unless the longitudinal  
($B_z$) field component becomes dynamically significant in these zones. 
 
\smallskip 
 
\item{2)} Concentric field structure in the region affected by instability is not preserved.  Once  
$m=1$ instability becomes well-established, it will probably wash out significant {\it mean}  
field gradients in the affected region.  Depending on the efficiency of reconnection, the 
reorganized field may retain considerable small-scale structure, with significant {\it local} 
gradients.  In particular, if the field structure evolves through  
axisymmetric expansion of a medium with an initially weak field --- a plausible scenario in jets  
and in the Crab Nebula --- then the field gradient may never become strong enough ($\beta >  
2/ (3\gamma)$, $d \log B / d \log r < - 1/2$) to shut off the instability.  The reorganized field  
might remain predominantly aligned with the toroidal plane, making it {\it appear} (from   
polarization measurements) that the field remains well-organized. 
 
\smallskip 
 
Since the Z--pinch is unstable down to very short wavelengths (and the presence of a small  
$B_z$ may even enhance instability depending on the helicity of the mode), inclusion of a finite  
resistivity may lead to some dissipation of magnetic energy and nonadiabatic behavior even in  
the linear regime, as well as to damping of the instability on the smallest scales.   
 
 \bigskip  
  
\goodbreak 
 
\centerline{\sechead 5. APPLICATIONS}  
  
\medskip  
 
As argued in the Introduction, the magnetic field structures of astrophysical jets and the  
interiors of plerions tend to become dominated by concentric loops of toroidal field.  Even if  
magnetic stresses dominate over pressure forces at large cylindrical radii, there must be a range  
of radii within which the condition $\beta > 2/(3\gamma) $ holds.  This region is susceptible to  
the $m=1$ instability discussed in this paper.  As loops of field in the  unstable zone are  
dislodged from concentric alignment by the instability, field loops from outer, stable zones may  
shrink to take their place.  But these too will then be subject to the $m = 1$ instability.  Thus  
we claim that instability will substantially destroy the organized field structures of astrophysical  
Z--pinches, provided that the unstable modes have time to grow into the nonlinear regime.    
Below we estimate the growth criterion for generic conditions in jets, and for the best-studied  
plerion, the Crab Nebula.  In each case, operation of the instability has important consequences  
for our understanding of physical conditions in these objects. 
 
\medskip 
 
\centerline{5.1. {\it Jets} }  
  
\medskip  
 
The growth times evaluated in \S~4 are measured in the frame of the material moving down  
the jet. Thus, if we examine a region of a jet at distance $R$ from the source, the instability will  
grow provided that $\Gamma t_g < R /v $, where $v$ is the jet flow speed and the bulk  
Lorentz factor, $\Gamma = (1 - v^2 / c^2)^{-1/2} $, corrects for time dilation.  If $r$ is a  
characteristic transverse scale across the jet (i.e., the jet's width), then the condition for growth  
of the instability in a nonrelativistic, supersonic jet is 
$$ { r \over R} < {v_A  \over v } = {1 \over M_A },  \eqno(5.1) $$ 
where $M_A$ is the Alfv\'en Mach number.  In a relativistic jet, it is convenient to express the  
growth criterion in terms of the ratio of Poynting flux to kinetic energy flux, the magnetization  
parameter (Michel 1969) 
$$ \sigma \equiv  {B^2 \over 4 \pi \varepsilon } , \eqno(5.2) $$ 
where $B$ and $\varepsilon$ are measured in the fluid frame.  The condition for growth of  
instability in a relativistic jet is then 
$$ { r \over R} < { \sigma^{1/2} \over \Gamma } .  \eqno(5.3)  $$ 
Since most jets are observed to expand laterally as they propagate away from the source, the  
ratio $r / R$ can be identified roughly with the rate of expansion of the jet. If the jet's interior is  
in causal contact with its surroundings --- e.g., if it is pressure-confined --- then the rate of  
expansion cannot exceed the Mach angle, i.e., 
$$ { r \over R} <  { 1 \over \Gamma } \left(  { p \over \varepsilon} \right)^{1/2}   
{ c \over v} \eqno(5.4)   
$$ 
for a region dominated by gas pressure, where we have neglected numerical factors of order  
unity to emphasize the basic scaling.  Observations of many jets indicate interactions with their  
environments, implying that eq.~(5.4) is probably satisfied (e.g., Bicknell \& Begelman 1996).   
 
Equation (5.4) alone is not sufficient to guarantee that the instability growth condition  
(eq.~[5.1] or [5.3]) is satisfied.  The upper limit on $r/R$ exceeds the instability growth limit by  
a factor $\sim \beta^{1/2} $, which is larger than one for a gas pressure-dominated jet.  Yet  
there are two conditions under which a jet flow will evolve toward satisfying the instability  
growth criterion with increasing distance from the source, even if it starts out stable under the  
criterion. First, a pressure-confined jet, propagating down a pressure gradient shallower than  
$r^{-2}$, will become increasingly collimated (Begelman, Blandford, \& Rees 1984). The ratio  
$r/ R$ decreases with $R$, while the right-hand side of the growth criterion remains constant  
(for matter-dominated jets) or increases (for relativistic flows with $p \gg \rho c^2$).  Second,  
the growth criterion will be satisfied automatically if the mean $\beta$ in the jet approaches  
unity.  This can occur even at constant $r/ R$, since the gas pressure in adiabatic expansion  
decreases more steeply with $r$ than does $B^2$. 
 
The conditions under which jets form are not well enough understood to make detailed  
predictions about the radius at which the instability will develop. We note that its onset may  
take place far from the jet's source and could lead to the dissipation of a large fraction of the  
magnetic energy into energetic particles.  There is observational evidence for the onset of  
dissipation at some distance from the source of AGN jets, in the form of synchrotron ``gaps''  
near the jets' bases (Bridle \& Perley 1984). An analysis of synchrotron emission along the jet  
in M87 also indicates that much of the particle acceleration takes place at 1--10 parsecs  
($10^3-10^4$ Schwarzschild radii) from the central black hole (Heinz \& Begelman 1997), i.e.,  
in a region quite distinct from the jet formation zone at $\ltorder 0.01$ pc (Junor \& Biretta  
1995).   
 
\medskip  
 
\centerline{5.2. {\it The Crab Nebula} }  
  
\medskip 
 
To a first-order approximation, the Crab Nebula can be described as a bubble inflated by a  
mixture of relativistic particles and toroidal magnetic field injected by the central pulsar in the 
form of a relativistic wind (Rees \& Gunn 1974; Kennel \& Coroniti 1984a; Begelman and Li 
1992). A standing reverse shock is inferred to lie about 10\% of the way from the pulsar to the 
edge of the synchrotron nebula, at the outer boundary of the ``underluminous" region 
surrounding the pulsar (Kennel \& Coroniti 1984a, and references therein) and just interior to 
the time-variable optical ``wisps" (Scargle 1969; Hester et al.~1995).  
 
Outside the shock, hydrostatic equilibrium is maintained by a balance between relativistic 
particle pressure and the combination of magnetic pressure and tension.  The shocked wind 
must decelerate from $\gtorder c/3$ just outside the shock to the nebular expansion speed of 
$\approx 1500$ km s$^{-1}$ at the bubble's edge.  Flux-freezing and gasdynamical arguments 
then lead to the following picture for the magnetic field structure.  In the inner regions the field 
strength increases linearly with radius, while the dominant particle pressure is approximately 
uniform. If the bubble extends beyond the radius at which the magnetic pressure becomes 
comparable to the particle pressure, then the structure would increasingly resemble that of a 
pure magnetic pinch, with $B$ declining as $r^{-1}$ and gas pressure becoming relatively 
unimportant.  
 
The inference that the magnetic field strength is an increasing function of radius in the inner 
nebula has been used to constrain the form in which energy is injected by the pulsar wind.  If 
the ratio of magnetic pressure to gas pressure were larger than a few percent just outside the 
reverse shock, then the outer nebula would be strongly pinched and therefore highly elongated, 
in contradiction to observations (Rees \& Gunn 1974; Begelman \& Li 1992).  Moreover, the 
reverse shock would then be pushed inward from its observationally inferred radius, perhaps 
collapsing altogether (Rees \& Gunn 1974; Emmering \& Chevalier 1987). Thus, in order for 
the Rees--Gunn model to satisfy boundary conditions at both the reverse shock and the nebular 
boundary, the magnetic pressure must not become significantly larger than the gas pressure 
anywhere in the nebula. Taking into account the amplification of the magnetic field due to 
compression at the shock front, this implies that the ratio of Poynting flux to kinetic energy flux 
($\sigma$) just inside the shock is no more than a few times $10^{-3}$. 
  
Begelman \& Li (1992) confirmed this qualitative behavior in an exact magnetohydrodynamic 
(MHD) model, and pointed out that the structure depends on cylindrical, not spherical, radius. 
Although local interactions between the relativistic fluid and the denser filaments of ejecta 
undoubtedly occur, and may partially be responsible for confining the nebula (Hester et 
al.~1996), the Rees--Gunn model seems to provide a robust description of the nebula's gross 
features.  Thus one can view the Crab Nebula as a confined, cylindrical structure with both 
magnetic and particle pressure varying with cylindrical radius. In other words, to an excellent 
approximation the interior of the Crab Nebula should resemble a Z--pinch. 
 
Indeed, the Rees--Gunn model depends crucially on the assumption that the magnetic field 
retains a concentric toroidal geometry as the nebula expands.  Is this realistic, given the 
instabilities to which Z--pinches are susceptible?  To address this question, consider the unique, 
two--parameter family of cylindrically symmetric MHD equilibria found by Begelman \& Li 
(1992).  These equilibria are characterized by the balance between particle pressure and the 
forces due to a toroidal magnetic field, and subject to the dual constraints of magnetic flux-
freezing and adiabatic flow along streamlines.  Although the specific entropy of the gas is 
allowed to vary arbitrarily from streamline to streamline, the equilibrium structure turns out to 
be completely independent of the flow pattern.  These models are ideally suited to describing 
the internal structure of the Crab Nebula and other plerions, at radii larger than the pulsar wind 
shock and in the limit that radiative losses have negligible effect on the equation of state. The 
models satisfy eq.~(4.7) exactly, with $\alpha_0 = 1$, i.e., 
$$  { d \ln B \over d\ln r } =  { \gamma\beta  - 2 \over \gamma\beta + 2} . \eqno(5.5) $$  
These equilibria are neutrally stable to $m=0$ modes, but are unstable to $m=1$ modes for  
$\beta > 1/2 $, where we have set the adiabatic index $\gamma = 4/3$ to describe the  
ultrarelativistic fluid that is presumably injected by the shocked pulsar wind.   
 
At the radius of the pulsar wind shock, $r_s$, the Rankine-Hugoniot conditions for a  
transverse MHD shock (Kennel \& Coroniti 1984) allow us to deduce the downstream $\beta- 
$parameter in terms of the magnetization parameter in the wind immediately upstream of the  
shock: 
$$ \beta_s = {4 \over 27 \sigma_s }.   \eqno(5.6)  $$ 
If we assume that $\beta_s$ is $\gtorder 1$ (but not necessarily as large as the values deduced  
by Rees \& Gunn and their successors), then flow just outside the shock is gas  
pressure-dominated. The postshock flow speed is then $v \approx (c/3) (r / r_s)^{-2}$, while  
$\beta \approx \beta_s (r / r_s)^{-2}$.  Instability sets in at $r_i$ where  
$$\int^{r_i}_{r_s}    { dr \over  v }  \sim (2 \beta)^{1/2} r_i /c$$  
(see eq.~[4.14]), giving  
$$ {r_i \over r_s } \sim \left[   1 +  (2 \beta_s )^{1/2} \right]^{1/3} .  \eqno(5.7)  $$ 
For the largest values of $\beta_s$ ($\approx 150$) derived using the Rees--Gunn model,  
eq.~(5.7) gives $r_i / r_s \ltorder 3$.   
 
Thus the interior of the Crab Nebula should be highly susceptible to the local $m = 1$  
instabilities described earlier.  If the instability reaches nonlinear amplitude as close to the pulsar  
wind shock as suggested by eq.~(5.7), then very little of the nebula should possess the  
concentric field structure inferred by Rees \& Gunn (1974).  Loops of toroidal field deposited  
by the pulsar wind at $r_s$ will not simply expand about a common axis, but will be dislodged  
and ``float away" before they expand very much at all.  Instead of a field geometry dominated  
by nested loops of field a parsec or more across (i.e., comparable to the size of the radio 
synchrotron nebula, $R_{\rm neb}$), we expect a chaotic structure with ``loops'' or field 
reversals common on scales $\sim r_s$ or less throughout the nebula (albeit possibly with a 
preferred field orientation --- see below).   Reconnection may be widespread, although we are 
unable to estimate its importance in dissipating magnetic energy.  
 
Destruction of the concentric field structure by $m=1$ instability is not inconsistent with the  
high degree of polarization and preferred polarization direction observed in the Crab Nebula  
(Schmidt, Angel, \& Beaver 1979; Michel et al.~1991).  Recall that the most unstable modes  
have radial wavelengths much longer than their wavelengths along the cylindrical axis (\S~4.2).   
Since unstable motions occur mainly in the $r-\phi$ plane (at least in the linear regime),  it  
seems plausible that $B_z$ remains small, even as $B_\phi$ and $B_r$ become randomized.  If  
the rotation axis of the pulsar lies not too far from the plane of the sky, the randomized field  
would still produce a high degree of synchrotron polarization, with a direction preferentially  
parallel to the projected rotation axis.  In fact, high resolution optical and radio observations of  
the nebula (Hickson \& van den Bergh 1990; Bietenholz \& Kronberg 1991), the latter  
corrected for foreground Faraday rotation, reveal a cellular structure, with a preferred direction  
overall but with fluctuations having a characteristic scale size of about $10^{\prime\prime}$ 
($= 0.1$ pc at the  assumed distance of 2 kpc).  This scale size is very similar to the inferred 
value of $r_s$, and supports our conjecture that the loops of toroidal field never grow beyond 
this scale before  
instability sets in. 
 
If our picture of a chaotic magnetic structure in the Crab is correct, then we must discard the  
dynamical argument used by Rees \& Gunn (1974) and their successors to estimate $\sigma$.  
To see this, it is useful to recall the ``winding number'' analogy introduced in Rees \& Gunn 
(see their eq.~[4]).  The magnetic energy inside the nebula satisfies  
$$ {\cal E}_{\rm mag} \propto {{\cal N}^2 \over  R_{\rm neb} } \left( \ell \over R_{\rm neb}  
\right)^2 ,   \eqno(5.8)  $$ 
where ${\cal N}$ is the number of turns made by the pulsar and $\ell$ is the mean size of a 
field loop.  In the Rees--Gunn model, $\ell \sim R_{\rm neb}$, since all of the loops expand  
concentrically. Rees \& Gunn (1974) therefore implicitly set this factor equal to one, and did  
not include it explicitly in their formula.  If the instability operates as we conjecture, then $\ell  
\sim r_s$ and $( \ell /  R_{\rm neb})^2 \sim 0.01 $.  To put this another way, the magnetic  
pressure would not increase with radius as $r^2$.  Therefore, to match the pressure and  
expansion rate of the nebula to its observed values, we can tolerate a value of $\sigma$ a  
hundred times larger than that inferred by Rees \& Gunn (1974).       
 
\bigskip 
 
\centerline{\sechead 6. CONCLUSIONS}  
  
\medskip  
 
I have analyzed the local stability of axisymmetric perfect MHD equilibria in which gas  
pressure balances the pressure and tension of a toroidal magnetic field.  This type of  
equilibrium  --- the ``Z--pinch'' of plasma physics literature --- is germane to plerions,  
like the Crab Nebula, and probably applies to many astrophysical jets as well.   The essential  
results of the stability analysis are not new --- they can be found in any number of elementary  
plasma physics or MHD textbooks, such as Bateman (1978) or Krall \& Trivelpiece (1973).  I  
present a complete derivation and discussion of the instability here because 1) the astrophysical  
applications require a relativistic treatment of the instability, which I have not found in the  
literature, and 2) I want to make the physics of these important instabilities more accessible to  
astrophysicists. 
 
Provided that they do indeed possess the assumed field geometry, jets and plerions are highly  
susceptible to internal kink instabilities, which probably destroy the concentric field structure.   
Unless jets possess a relatively coherent ``backbone" of longitudinal ($B_z$) field, without  
field reversals on scales much smaller than the jet width, they are unlikely to develop any  
significant degree of magnetic self-collimation.  However, the slow growth rate of the  
instability in a weak magnetic field implies that a well-ordered toroidal field structure could  
persist for a significant distance along the jet, depending on how tightly collimated the jet is.  
Dissipation associated with the eventual onset of instability could provide a source of energy  
for particle acceleration at otherwise unremarkable regions of a jet far from its source. 
 
In the Crab Nebula we expect the instability to develop very close to the pulsar wind shock, at  
distances of only $\sim 0.1$ pc from the pulsar. Disruption of the concentric field structure  
would imply that the mean field strength is not amplified much by the expansion of the flow 
between the wind shock and the outer edge of the nebula.  The problem of confining the 
concentric toroidal field, first noted by Rees \& Gunn (1974), goes away, and with it goes their 
deduction that $\sigma$ --- the ratio of Poynting flux to kinetic energy flux --- in the pulsar's 
wind must be $\ll 1$. 
 
The low value of $\sigma$ derived from the Rees--Gunn model has puzzled theorists for years.   
Perfect MHD models of pulsar winds predict that $\sigma$ cannot become much smaller than 
one (Michel 1969; Kennel et al.~1983; Arons 1992), unless highly contrived boundary 
conditions are satisfied (Begelman \& Li 1994; Tomimatsu 1994).  Attempts to explain the 
decline in $\sigma$ through dissipative effects have led to some ingenious but controversial 
proposals (Coroniti 1990; Michel 1994; Melatos \& Melrose 1996). While analyses of the 
nebula's synchrotron spectrum and emissivity (Kennel \& Coroniti 1984b; Hoshino et al.~1992) 
have lent support to a low-$\sigma$ wind, it has always been the dynamical argument that has 
seemed most incontrovertible. The proposal we have made --- that the  
Rees--Gunn field geometry is physically unrealistic --- does not allow us to determine the value 
of $\sigma$ in the Crab pulsar wind.  However, it does eliminate the principal dynamical 
argument for concluding that $\sigma$ is small.  Spectral arguments nonwithstanding , it would 
seem worthwhile to reconsider whether $\sigma$ might be as large as $\sim O(1)$, after all. 
 
\bigskip 
 
I am grateful to Martin Rees for his valuable insights during the three-year gestation period of  
this paper, and for encouraging me to finish it. Ellen Zweibel critiqued an early draft and 
supplied advice on the plasma physics literature. This work has been supported in part by  
National Science Foundation grants AST-9120599 and AST-9529175. 
 
\bigskip
\vfill\eject
 
\centerline{\sechead REFERENCES}  
  
\def\ref{ \parindent=0pt  
\hangindent=20pt  
\hangafter=1 \smallskip}  
  
\ref Arons, J. 1992, in The Magnetospheric Structure and Emission Mechanisms of Radio   
Pulsars, IAU Colloq.~128, ed. J.~A.~Gil, T.~H.~Hankins, \& J.~M.~Rankin (Zielona G\'ora:   
Pedagogical Univ.~Press), p.~56 
 
\ref Bateman, G. 1978, MHD Instabilities (Cambridge: MIT  Press)  
  
\ref Begelman, M.~C., Blandford, R.~D., \& Rees, M.~J. 1984, Rev.~Mod.~Phys., 56, 255  
  
\ref Begelman, M.~C., \& Li, Z. 1992, ApJ, 397, 187  
  
\ref Begelman, M.~C., \& Li, Z. 1994, ApJ, 426, 269  
  
\ref Benford, G. 1978, MNRAS, 183, 29  
 
\ref Bicknell, G.~V., \& Begelman, M.~C. 1996, ApJ, 467, 597 

\ref Bicknell, G.~V., \& Henriksen, R.~N. 1980, ApLett, 21, 29
 
\ref Bietenholz, M.~F., \& Kronberg, P.~P. 1991, ApJ, 368, 231 
 
\ref Bridle, A.~H., \& Perley, R.~A. 1984, ARA\&A, 22, 319 

\ref Chan, K.~L., \& Henriksen, R.~N. 1980, ApJ, 241, 534
 
\ref Chen, F.~F. 1974, Introduction to Plasma Physics (New York: Plenum) 
  
\ref Coroniti, F.~V. 1990, ApJ, 349, 538  

\ref Eichler, D. 1993, ApJ, 419, 111
  
\ref Emmering, R.~T., \& Chevalier, R.~A. 1987, ApJ, 321, 334  
  
\ref Freidberg, J.~P. 1982, Rev.~Mod.~Phys., 54, 801  
  
\ref Heinz, S., \& Begelman, M.~C. 1997, ApJ, in press 
 
\ref Hester, J.~J., et al. 1995, ApJ, 448, 240 
 
\ref Hester, J.~J., et al. 1996, ApJ, 456, 225 
 
\ref Hickson, P., \& van den Bergh, S. 1990, ApJ, 365, 224 
 
\ref Jackson, J.~D. 1975, Classical Electrodynamics (New York: John Wiley \& Sons) 
 
\ref Junor, W., \& Biretta, J.~A. 1995, AJ, 109, 500 
 
\ref Hoshino, M., Arons, J., Gallant, Y.~A., \& Langdon, A.~B. 1992, ApJ, 390, 454 
 
\ref Kadomtsev, B.~B. 1966, in Reviews of Plasma Physics, 2, ed. M.~A.~Leontovich (New   
York: Consultants Bureau), p.~166  
  
\ref Kennel, C.~F., \& Coroniti, F.~V. 1984a, ApJ, 283, 694  
 
\ref Kennel, C.~F., \& Coroniti, F.~V. 1984b, ApJ, 283, 710 
  
\ref Kennel, C.~F., Fujimura, F.~S., \& Okamoto, I. 1983,  
J.~Astrophys.~Geophys.~Fluid.~Dyn., 26, 147  
  
\ref K\"onigl, A., \& Choudhuri, A.~R. 1985, ApJ, 289, 173 
 
\ref Krall, N.~A., \& Trivelpiece, A.~W. 1973, Principles of Plasma Physics (New York:  
McGraw-Hill)  
  
\ref Kruskal, M., \& Schwarzschild, M. 1954, Proc.~Roy.~Soc., A223, 348   
  
\ref Melatos, A., \& Melrose, D.~B. 1996, MNRAS, 279, 1168  
  
\ref Michel, F.~C. 1969, ApJ, 158, 727  
  
\ref Michel, F.~C. 1994, ApJ, 431, 397  
 
\ref  Michel, F.~C., Scowen, P.~A., Dufour, R.~J., \& Hester,  J.~J. 1991,  ApJ, 368, 463  
 
\ref  Rees, M.~J., \& Gunn, J.~E. 1974, MNRAS, 167, 1  
 
\ref  Scargle, J.~D. 1969, ApJ, 156, 401 
 
\ref  Schmidt, G.~R., Angel, J.~R.~P., \& Beaver, E.~A. 1979,  
ApJ, 227, 106  
 
\ref Tomimatsu, A. 1994, PASJ, 46, 123  
  
\ref Weinberg, S. 1972, Gravitation and Cosmology: Principles and Applications of the General   
Theory of Relativity (New York: John Wiley \& Sons)  
  
\bye